\documentclass[twocolumn,aps,superscriptaddress,nofootinbib]{revtex4}

\usepackage{amsmath,amssymb}
\usepackage{graphicx}
\usepackage{dcolumn}
\usepackage{bm}
\usepackage[usenames]{color}
\usepackage[breaklinks=true,colorlinks=true,linkcolor=blue,urlcolor=blue,citecolor=blue]{hyperref}
\usepackage{ulem}
\usepackage{amsmath}
\usepackage{float}
\usepackage{ulem}

\usepackage{comment}

\begin{document}

\title{Translocation of an Active Polymer into a Circular Cavity}
 
\author{Amir Rezaie-Dereshgi}
\affiliation{School of Nano Science, Institute for Research in Fundamental Sciences (IPM), 
19395-5531, Tehran, Iran.}
\affiliation{Department of Physics, Institute for Advanced Studies in
Basic Sciences (IASBS), 45137-66731, Zanjan, Iran.}

\author{Hamidreza Khalilian}
\affiliation{School of Nano Science, Institute for Research in Fundamental Sciences (IPM), 
19395-5531, Tehran, Iran.}

\author{Jalal Sarabadani}
\email{jalal@ipm.ir}
\affiliation{School of Nano Science, Institute for Research in Fundamental Sciences (IPM), 
19395-5531, Tehran, Iran.}

\date{\today}

\begin{abstract}
Translocation dynamics of an active semi-flexible polymer through a nano-pore into a rigid two dimensional circular cavity, and the polymer packing dynamics have been studied by using Langevin dynamics (LD) simulations. The results show that the force exponent $\beta$, for regime of small cavity radius, i.e. $R \ll R_{\textrm{g}}$, where $R_{\textrm{g}}$ is the gyration radius of the passive semi-flexible polymer in  two dimensional free space, is $\beta=-1$, while for large values of $R \gg R_{\textrm{g}}$ the asymptotic value of the force exponent is $\beta \approx -0.92$. The force exponent is defined by the scaling form of the average translocation time $\langle \tau \rangle \propto F_{\textrm{sp}}^{\beta}$, where $F_{\textrm{sp}}$ is the self-propelling force. Moreover, using the definition of the turning number for the polymer inside the cavity, it has been found that at the end of translocation process for small value of $R$ and in the strong force limit the polymer configuration is more regular than the case in which the value of $R$ is large or the force is weak. 
\end{abstract}

\maketitle


\section{Introduction}  \label{secl_INTRO}

Polymer translocation through a nano-pore, which makes a bridge between physics and  biology, has attracted a lot of attention since the seminal experimental~\cite{bezrukov1994counting, kasianowicz1996characterization} and theoretical works~\cite{sung1996polymer} in about three decades ago. Translocation of a DNA and RNA through a nano-pore on a nuclear membrane, virus genome injection and transportation of a protein chain across a pore on a membrane of a cell, are some important examples of polymer translocation in biology \cite{muthukumar2016polymer}. Moreover, polymer translocation through a nano-pore has been used to develop some nano-technological tools to manipulate a single molecule in experiments, and a long cylindrical nano-pore setup has been used to visualize the genomic information by investigation the DNA spatial conformations~\!\cite{Tegenfeldt_book,Tegenfeldt_CSR2010,Tegenfeldt_ABC2004,Tegenfeldt_PNAS2005,Tegenfeldt_PNAS2004}.
Such a diverse and extensive subject, from essential role of polymer translocation in the cell function to its experimental applications has attracted people to study this phenomenon. 
An important factor that affects the translocation dynamics is the geometry of the {\it trans} side, i.e. the available space for the polymer in the {\it trans} side, which can be semi-space~\!\cite{sung1996polymer, bhattacharya2009scaling, luo2007heteropolymer, luo2008dynamics, rowghanian2011force, saito2011dynamical, saito2012process, sakaue2007nonequilibrium, sakaue2010sucking, cohen2011active, sarabadani2014iso, sarabadani2015theory, sarabadani2017driven, sarabadani2018theory, sarabadani2018dynamics, menais2018polymer, sarabadani2020pulling, dubbeldam2007polymer, gauthier2009nondriven, huopaniemi2006langevin, luo2007influence, luo2008dynamical, milchev2004polymer, palyulin2014polymer}, confined in one dimension~\!\cite{luo2010polymerpre, luo2010polymerjcp,  sarabadani2022driven}, or confined in two dimensions~\cite{ali2006polymer, das2019dynamics, goldfeld2009packaging, purohit2003mechanics, sakaue2007semiflexible, smith2001bacteriophage, sun2018theoretical, zhang2014polymer}. The latter case corresponds to the two dimensional model for packing of a DNA or RNA of a virus inside a capsid.

\begin{figure}[b]
	\begin{minipage}{0.5\textwidth}
    \begin{center}
        \includegraphics[width=0.85\textwidth]{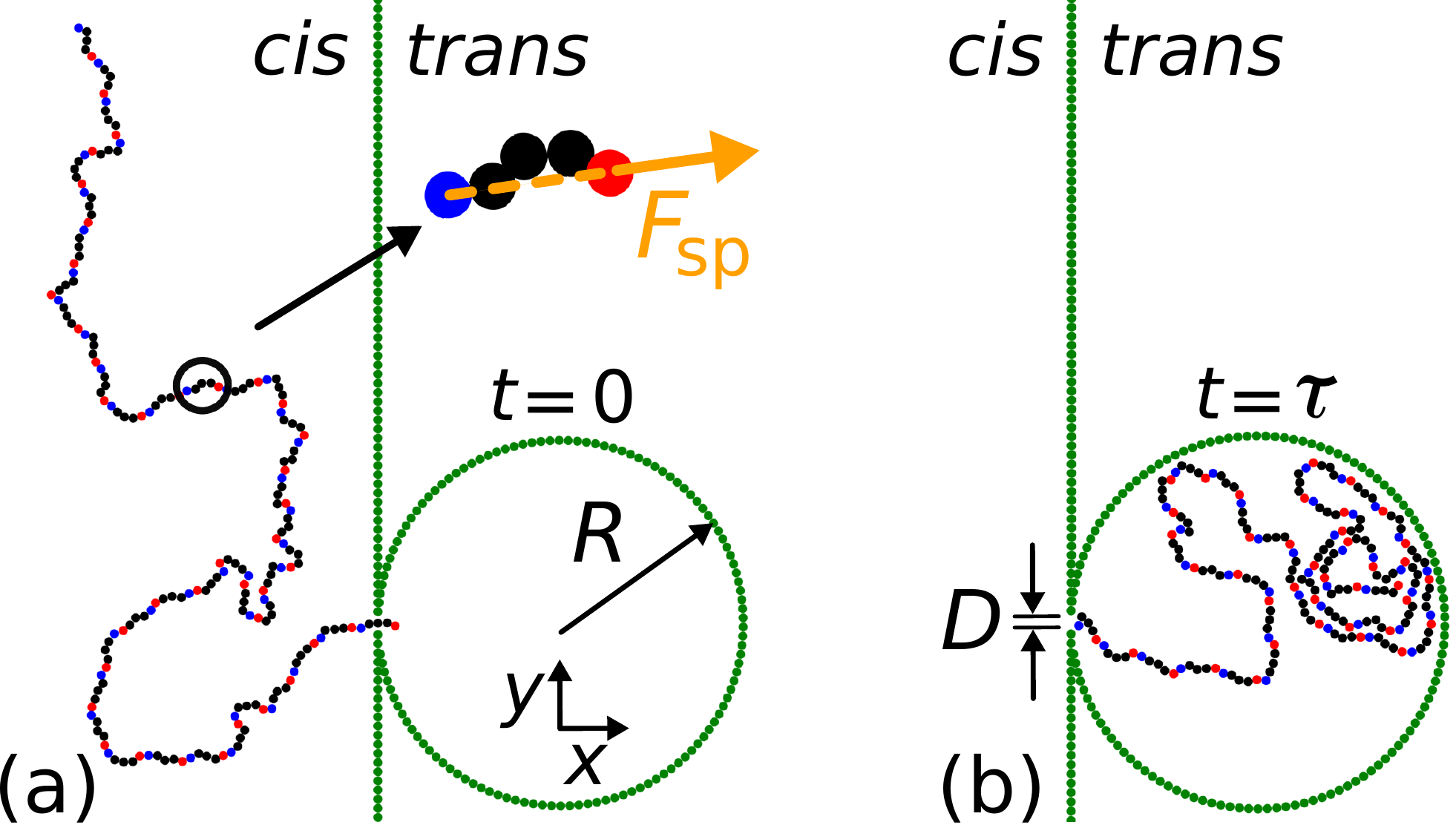}
    \end{center}
    \end{minipage}
\caption{Schematic of an active polymer translocation through a nano-pore into a rigid two dimensional circular cavity with radius of $R=20$, (a) at $t=0$ (beginning of the translocation process), and (b) at $t = \tau$ (end of the process). The rigid wall and two dimensional circular rigid cavity, which are depicted by green beads, are located in the $x-y$ plane. The wall, which is parallel to the $y$ axis, is located at $x=0$. The circular cavity whose center is at $(R,0)$ has been tangentially attached to the wall. The center of the nano-pore with width of $D = 1.5$ is located at $(0,0)$. Magnified picture shows the tail (in blue) and head (in red) monomers belonging to a segment. At each time step the SP force acts on the red head monomer in the direction from the tail to the head bead. The contour length of the polymer is $N=200$ and its persistence length is $l_{p}=5$. For this figure the magnitude of the SP force is $F_{\textrm{sp}}=2$.}
\label{fig_snapshot}    
\end{figure}

To pass through a nano-pore the conformation of a polymer has to be changed and due to decrease in the available polymer conformations during the translocation process, the polymer feels an entropic barrier. To have a successful translocation event, the polymer must overcome this entropic barrier by using different strategies such as external fields~\cite{sigalov2008detection, meller2001voltage}, flow-induced~\cite{auger2014zero, chen2021dynamics, ding2021flow, sakaue2005flow, zheng2018revisiting}, chaperon-assisted~\cite{khalilian2021polymer, abdolvahab2011sequence, ambjornsson2005directed, chen2013dynamics, emamyari2017polymer, gopinathan2007polymer, taheri2018granular, yu2011chaperone, zandi2003drives}, and pulling techniques like atomic force microscope (AFM), optical and magnetic tweezers~\cite{ritort2006single, keyser2006direct}. 
%
From a theoretical point of view, driven translocation of a polymer through a nano-pore can be divided into two categories pore-driven~\!\cite{bhattacharya2009scaling, luo2008dynamics, rowghanian2011force, saito2011dynamical, saito2012process, sakaue2007nonequilibrium, sakaue2010sucking, sakaue2016dynamics, sarabadani2014iso, sarabadani2015theory, sarabadani2018theory, sarabadani2017driven, cohen2011active, zhang2014polymer} and end-pulled~\!\cite{menais2018polymer, huopaniemi2007polymer, menais2016polymer, sarabadani2018dynamics, sarabadani2018theory, sarabadani2020pulling}.
In the present paper, instead of the explicit external driving force, a new strategy is considered in which the self-propulsion of the polymer plays an effective driving force role that facilitates the translocation.

Active polymers and filamentous structures play a crucial role in the biological systems and have various  experimental applications~\!\cite{phillips2012physical, howard2001_book}. Accordingly, dynamical and  conformational properties of active polymers have been studied theoretically (through both the analytical as well as the simulation models)~\!\cite{anand2018_pre, Gompper_tangentional_activity, eisenstecken2017_jcp, eisenstecken2022_jcp, isele2015_sm, laskar2017_njp, peterson2020_JSM-TE, sarkar2016_pre} and experimentally~\!\cite{dreyfus2005_Nature, hill2014_ACS-Nano, nishiguchi2018_njp, schaller2010_Nature}. As a manifestation of active polymer applications, the large group of biological and artificial micro-swimmers, by using filamentous and hairy like organelles, such as flagella and cilia as the active bio-polymers, produce a net locomotion in the low Reynolds regimes~\!\cite{phillips2012physical, howard2001_book, brennen1977_ARFM, elgeti2015_RPP, bechinger2016_RMP}. 


In this paper, our purpose is to study the translocation of an active polymer through a nano-pore into a two dimensional circular cavity by overcome the entropic barrier. 
The present paper is organized as follows: In Sec.~\ref{secl_MM} the Langevin dynamics (LD) simulation method is explained in detail and it is introduced how to model a coarse grained self-propelled semi-flexible polymer as well as a rigid two dimensional cavity. 
Then in Sec.~\ref{secl_RES} the results are presented. Finally, summary and conclusions are discussed in Sec.~\ref{secl_SC}. 


\section{Models and Methods} 
\label{secl_MM}

\subsection{Simulation model}
\label{subl_sm}

This paper is devoted to present our study on the translocation of a self-propelled semi-flexible polymer, consisting of $N$ identical monomers, through a nano-pore into a two dimensional circular cavity using the Langevin dynamics (LD) simulation method by means of the LAMMPS package~\cite{LAMMPS}. The configurations of the system at the beginning of the translocation process at time $t=0$, and at the end of the translocation process at time $t = \tau$ ($\tau$ is the translocation time) have been illustrated in Fig.~\!\ref{fig_snapshot}(a) and (b), respectively. As shown a rigid wall, which is located in the $x-y$ plane parallel to the $y$ axis at $x=0$, separates the system into {\it cis} and {\it trans} sides. 
The center of the nano-pore, whose width is $D = 1.5$, is located at $(0,0)$ and is at the intersection of the wall and cavity boarder. Moreover, the center of a rigid cavity with radius of $R$ is at $(R,0)$ and the cavity has been attached tangentially to the wall. It should be mentioned that the wall and the cavity have been made by similar beads with size $\sigma$. During the translocation process a self-propelled polymer translocates through a nano-scale pore from the {\it cis} side ($x<0$), into the cavity located in the {\it trans} side ($x>0$).


\subsection{Langevin dynamics simulation}
\label{subl_lds}

In our simulations, a semi-flexible self-avoiding polymer composed of $N$ monomers has been studied as a bead-spring chain. The bonded interactions are the sum of the Weeks-Chandler-Anderson (WCA) and finitely extensible nonlinear elastic (FENE) potentials. The repulsive WCA potential is 
\begin{equation}
U_{\textrm{WCA}}(r)=
\begin{cases}
U_{\textrm{LJ}}(r)-U_{\textrm{LJ}}(r_{\textrm{c}}), & \text{if $r \leq r_{\textrm{c}}$};\\
0, & \text{if $r > r_{\textrm{c}}$,}
\end{cases}
\label{eq_ljs}
\end{equation}
where $r$ is the distance between two monomers, $r_{\textrm{c}} = 2 ^ {1/6}$ is the cut-off radius and $U_{\textrm{LJ}}(r) = 4 \varepsilon \big[ \big( \sigma/r\big)^{12} - \big( \sigma/r \big)^6 \big]$ is the Lennard-Jones (LJ) potential with  $\varepsilon$ and $\sigma$ as the depth of the potential and the monomer size, respectively. The consecutive bonded monomers are connected to each other by the FENE potential 
\begin{equation}
 U_{\textrm{FENE}}(r) = -\frac{1}{2}k R^{2}_{0}\ln \bigg[ 1 - \bigg( \frac{r}{R_{0}} \bigg)^{2} \bigg],
 \label{eq_fene}
\end{equation}
where $R_0$ and $k$ are the maximum allowed distance between the monomers and the spring constant, respectively. The stiffness of the chain comes from the bending potential
\begin{equation}
 U_{\textrm{b}}(\theta) = k_{\textrm{b}} [ 1 + \cos(\theta) ],
 \label{eq_bending}
\end{equation}
where $k_{\textrm{b}}$ and $\theta$ denote the bending energy coefficient and angle between two consecutive bonds, respectively. In the two dimensional system, the persistence length is given by $l_{\textrm{p}}= 2 k_{\textrm{b}} / (k_{\textrm{B}} T)$, where $k_{\textrm{B}}$ and $T$ are the Boltzmann constant and temperature, respectively. Moreover, the non-bonded interactions between monomers and the wall as well as the cavity are WCA.

To activate the polymer, the contour length of the polymer is divided to $n=N/l_{\textrm{p}}$ segments each with the length of $l_{\textrm{p}}$. In the system under consideration, the contour length and the persistence length of the polymer are constant and are $N=200$ and $l_{\textrm{p}}=5$, respectively. Therefore, the polymer consists of $n=40$ segments. As can be seen from the magnified segment of the polymer in Fig.~\!\ref{fig_snapshot}(a), the tail and head monomers are in blue and red colors, respectively, and three beads in the body of each segment are in black color. At each time step, the direction of the self-propelling (SP) force on each segment is updated according to the updated positions of the monomers, and a SP force $F_{\textrm{sp}}$ acts on the head of each segment in the direction of the tail bead toward the head bead of that segment.

As mentioned above the LD simulation method has been employed to simulate the dynamics of the system. Therefore, the equation of motion of the $i$th monomer is written as
\begin{equation}
M\ddot{\vec{r}}_{i} = - \vec{\nabla}U_{i} - \eta \dot{\vec{r}}_{i} + F_{\textrm{sp}}\delta_{i\textrm{h}} \bold{\hat{e}}_{\textrm{h}} + \vec{\xi}_{i} (t),
\label{eq_LD}
\end{equation}
where, $\vec{r}_{i}$ is the position vector of the $i$th monomer and $U_{i}$ is the sum of all interaction potentials experienced by the $i$th monomer. The $M$, $\eta$ and $F_{\textrm{sp}}$ denote the monomer mass, friction coefficient of the solvent, and the magnitude of the SP force, respectively. $\delta_{i\textrm{h}}$ is the Kronecker delta function for ensuring that the SP force $F_{\textrm{sp}}$ just acts on the head of each segment and $\bold{\hat{e}}_{\textrm{h}}$ represents the unit vector at the direction of the tail monomer to the head monomer in each segment at the corresponding time. $\vec{\xi}_{i}(t)$ is the thermal white noise at time $t$ with the average of $\langle \vec{\xi}_{i}(t) \rangle = \vec{0} $ and $ \langle \vec{\xi}_{i}(t) \cdot \vec{\xi}_{j}(t') \rangle = 4 \eta k_{\textrm{B}} T \delta_{ij} \delta(t - t') $, in which $\delta_{ij}$ and $\delta(t - t')$ are Kronecker and Dirac delta functions, respectively.

The $M$, $\sigma$ and $\varepsilon$ have been used as the units of mass, length and energy, respectively. The mass of each bead (including the monomers and also wall and cavity beads) is $M=1$ and the value of its diameter is $\sigma=1$. In addition, the values of other parameters are $\varepsilon=1$, $\eta=0.7$, $k_{\textrm{B}}T=1.2$, $k=30$ and $R_{0}=1.5$, and the value of the bending energy coefficient has been chosen as $k_{\textrm{b}}=3$ which is equivalent to the persistence length of $l_{p}=5$ in two dimension.
The size of the simulation box in the $x$ and $y$ directions are $L_{x}=L_{y}=420$ and there is a  periodic boundary condition in the $y$ direction on the {\it cis} side.

The initial configuration of the system is the straight polymer chain that its first two monomers are inside the cavity while its third monomer is fixed at the center of the nano-pore and the rest of the polymer beads are in the {\it cis} side. The LD simulations contain two stages. In the first stage, while the third polymer bead is fixed at the nano-pore the polymer is well equilibrated during the time interval of $t_{\textrm{eq}}=2\times10^{4}$ (in LJ unites). During the equilibration stage the polymer is passive. After the equilibration stage the fixed monomer at the nano-pore is released and simultaneously the SP forces are switched on and act on the heads of segments, and the second stage that is the actual translocation process of the polymer into the cavity is started.

\section{Results} 
\label{secl_RES}

In this section, the results of the translocation of a semi-flexible polymer composed of $N=200$ monomers, with persistence length $l_{\textrm{p}}=5$, into a circular cavity with various values of radius $R=10$, $12$, $15$, $20$, $30$, $40$, and $50$ are considered. The chosen values for the SP force are $F_{\textrm{sp}}=1.2$, 2, 5, 10, and 15. To find the mean values of different quantities, the averages have been performed over 1000 uncorrelated trajectories.

\subsection{Translocation time}
\label{subl_tt}

\begin{figure}[t]
	\begin{minipage}{0.5\textwidth}
    \begin{center}
    		\hspace{-0.6cm}
        \includegraphics[width=0.8\textwidth]{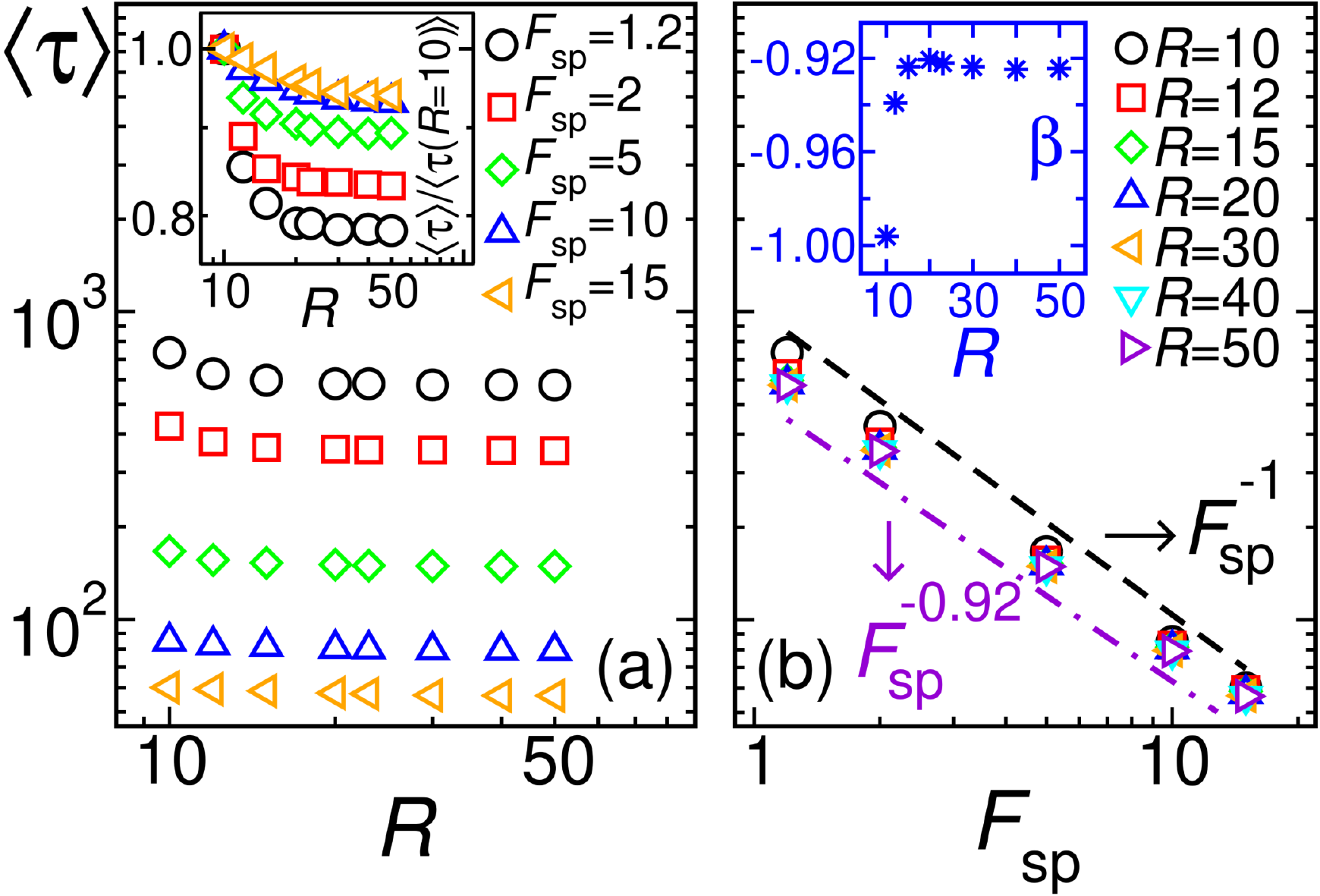}
    \end{center}
    \end{minipage} 
\caption{(a) The average of the translocation time $\langle \tau \rangle$ as a function of the cavity radius $R$, for different values of the SP force $F_{\textrm{sp}}=1.2$ (black circles), 2 (red squares), 5 (green diamonds), 10 (blue triangles up) and 15 (orange triangles left). The inset shows the normalized average translocation time $\langle \tau \rangle / \langle \tau (R=10) \rangle$ as a function of $R$. 
(b) The $\langle \tau \rangle$ as a function of the SP force $F_{\textrm{sp}}$ for various values of $R=10$ (black circles), 12 (red squares), 15 (green diamonds), 20 (blue triangles up), 30 (orange triangles left), 40 (cyan triangles down) and 50 (violet triangles right). The black dashed and violet dashed-dotted lines are fit to the data with $R=10$ and 50, respectively, but shifted for better visibility. The inset presents the force exponent $\beta$ of the scaling form $ \langle \tau \rangle  \propto F_{\textrm{sp}}^{\beta}$ as a function of $R$. 
}
\label{fig_translocation_time}
\end{figure}

The most important quantity, that characterizes the translocation dynamics both in the theory and experiment, is the translocation time $\tau$. The $\tau$ is the time that takes for the whole polymer to translocate into the {\it trans} side, i.e. into the cavity for the present system under consideration (see Fig.~\!\ref{fig_snapshot}). 
In Fig.~\!\ref{fig_translocation_time}(a) the average translocation time $\langle \tau \rangle$ has been plotted as a function of the cavity radius $R$, for various values of the SP force $F_{\textrm{sp}}=1.2$ (black circles), 2 (red squares), 5 (green diamonds), 10 (blue triangles up), and 15 (orange triangles left). In the inset the normalized average translocation time $\langle \tau \rangle / \langle \tau (R=10) \rangle $ has been plotted as a function of $R$. As can be seen $\langle \tau \rangle$ decreases monotonically as $R$ increases. Inset shows that the decreasing of $\langle \tau \rangle$ is more pronounced at weak force limit.
Panel (b) presents $\langle \tau \rangle$ as a function of $F_{\textrm{sp}}$ for different values of the cavity radius $R=10$ (black circles), 12 (red squares), 15 (green diamonds), 20 (blue triangles up), 30 (orange triangles left), 40 (cyan triangles down) and 50 (violet triangles right). The black dashed and violet dashed-dotted lines are fitting to the data for $R=10$ and 50, respectively, but shifted for better visibility. The inset in panel (b) shows the force exponent $\beta$, that is defined as $\langle \tau \rangle \propto F_{\textrm{sp}}^{\beta}$, as a function of the circular cavity radius $R$. As can be seen the value of $\beta$ approaches $-1$ as long as the radius of the circular cavity is much less than gyration radius of the passive semi-flexible polymer in free space in two dimensions $R_{\textrm{g}} \approx 23$, i.e. $R \ll R_{\textrm{g}}$. Then, $\beta$ increases rapidly by increasing the value of $R$ in the region $R \leq R_{\textrm{g}}$ and in the region $R \geq R_{\textrm{g}}$, it takes its asymptotic value $\beta \approx -0.92$. 

\begin{figure*}[t]
	\begin{minipage}{1.0\textwidth}
    \begin{center}
        \includegraphics[width=0.8\textwidth]{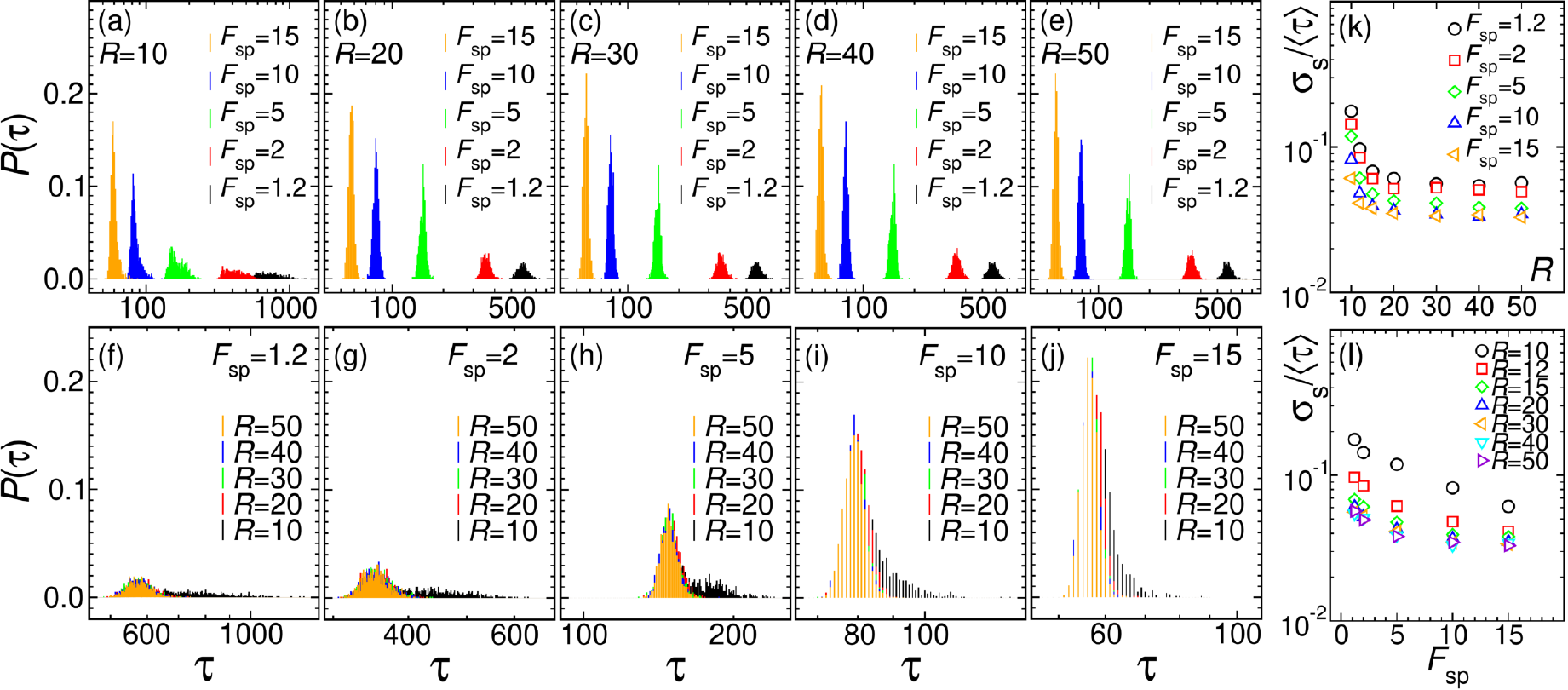}
    \end{center}
    \end{minipage} 
\caption{(a) Probability distribution function of the translocation times $P(\tau)$ as a function of translocation time $\tau$, for fixed value of the cavity radius $R=10$ and various values of the SP force $F_{\textrm{sp}}=15$ (orange bars), 10 (blue bars), 5 (green bars), 2 (red bars), and 1.2 (black bars). 
Panels (b), (c), (d) and (e) are the same as panel (a) but for different values of $R=20$, 30, 40 and 50, respectively. 
(f) The $P(\tau)$ as a function of $\tau$, for fixed value of the SP force $F_{\textrm{sp}}=1.2$ and various values of $R=50$ (orange bars), 40 (blue bars), 30 (green bars), 20 (red bars) and 10 (black bars). 
Panels (g), (h), (i) and (j) are the same as panel (f) but for different values of $F_{\textrm{sp}}=2$, 5, 10 and 15, respectively.
(k) Normalized value of the standard deviation $\sigma_{\textrm{s}} / \langle \tau \rangle $ as a function of $R$, for various values of $F_{\textrm{sp}}=1.2$ (black circles), 2 (red squares), 5 (green diamonds), 10 (blue triangles up) and 15 (orange triangles left). 
(l) The $\sigma_{\textrm{s}} / \langle \tau \rangle $ as a function of $F_{\textrm{sp}}$ for different values of $R=10$ (black circles), 12 (red squares), 15 (green diamonds), 20 (blue triangles up), 30 (orange triangles left), 40 (cyan triangles down) and 50 (violet triangles right). 
}
\label{fig_histogram_translocation_time}
\end{figure*}

\begin{figure*}[t]
	\begin{minipage}{1.0\textwidth}
    \begin{center}
        \includegraphics[width=0.98\textwidth]{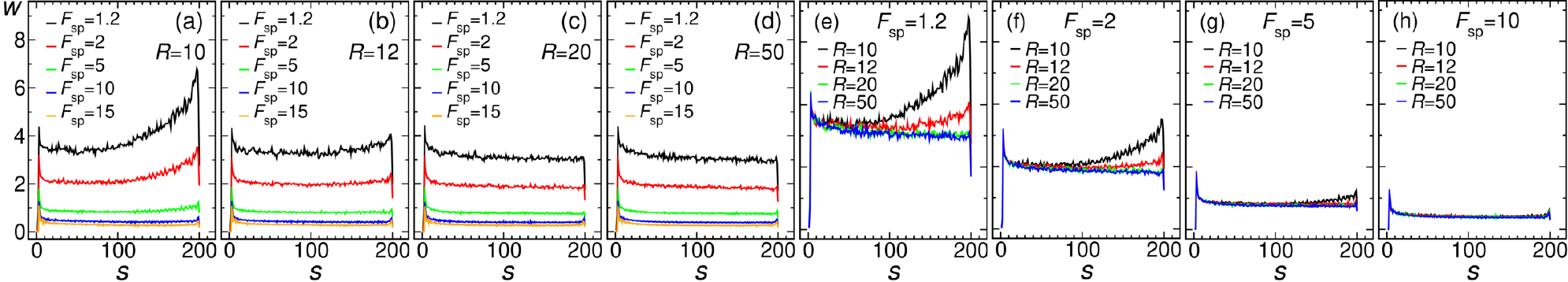}
    \end{center}
    \end{minipage} \hskip-1.33cm
\caption{(a) Waiting time $w$ as a function of the translocation coordinates $s$ for a fixed value of the cavity radius $R=10$ and various values of the SP force $F_{\textrm{sp}}=1.2$ (black line), 2 (red line), 5 (green line), 10 (blue line) and 15 (orange line). 
Panels (b), (c) and (d) are the same as panel (a) but for different values of $R=12$, 20 and 50, respectively. 
(e) The $w$ as a function of $s$ for a fixed value of $F_{\textrm{sp}}=1.2$ and various values of $R=10$ (black line), 12 (red line), 20 (green line) and 50 (blue line), respectively. 
Panels (f), (g) and (h) are the same as panel (e) but for different values of $F_{\textrm{sp}}=2$, 5 and 10, respectively.}
\label{fig_waiting_time}
\end{figure*}

In Fig.~\!\ref{fig_histogram_translocation_time}(a) the probability distribution function of translocation times $P(\tau)$ has been plotted as a function of translocation time $\tau$ for different values of the SP force $F_{\textrm{sp}}=1.2$ (black bars), 2 (red bars), 5 (green bars), 10 (blue bars), and 15 (orange bars), and fixed value of the cavity radius $R=10$. Panels (b), (c), (d) and (e) are the same as panel (a) but for cavity radii of $R=20$, 30, 40 and 50, respectively. As can be seen in each panel, for a given value of $R$, as the value of the SP force decreases the average value of translocation time $\langle \tau \rangle $ increases and the histograms get wider (note that the horizontal axis is in logarithmic scale). 
To illustrate the width of $P(\tau)$s of panels (a)--(e), in panel (k) the normalized standard deviation $\sigma_{\textrm{s}} / \langle \tau \rangle $ has been plotted as a function of the radius of the circular cavity $R$, for various values of the SP force $F_{\textrm{sp}}=1.2$ (black circles), 2 (red squares), 5 (green diamonds), 10 (blue triangles up), and 15 (orange triangles left). According to panels (a) to (e), $\sigma_{\textrm{s}} / \langle \tau \rangle $ decreases monotonically by increasing the value of $R$ and then for $R \gtrsim R_{\textrm{g}}$ it takes its asymptotic constant value for each value of the SP force.
Panel (f) shows $P(\tau)$ as a function of $\tau$ for different values of cavity radius $R=10$ (black bars), 20 (red bars), 30 (green bars), 40 (blue bars), and 50 (orange bars), for fixed value of SP force $F_{\textrm{sp}} = 1.2$. Panels (g), (h), (i) and (j) are the same as panel (f) but for different values of SP force $F_{\textrm{sp}} = 2$, 5, 10 and 15, respectively. As seen, for a given value of  $F_{\textrm{sp}}$ the probability distribution function of a translocation time $P(\tau)$ doesn't change significantly for $R \gtrsim R_{\textrm{g}}$. 
Panel (l) shows the same value as in panel (k) but as a function of the SP force, for different values of the radius of the cavity $R=10$ (black circles), 12 (red squares), 15 (green diamonds), 20 (blue triangles up), 30 (orange triangles left), 40 (cyan triangles down), 50 (violet triangles right). It can be seen that $\sigma_{\textrm{s}} / \langle \tau \rangle $ monotonically decreases by increasing the SP force, and for $R \gtrsim R_{\textrm{g}}$ the curves collapse on a master curve.
In summary, for a given value of $R$, the distributions of translocation times (and also the mean translocation times) significantly deviate from each other by increasing the value of $F_{\textrm{sp}}$ [see panels (a)--(e)]. On the other hand, for a given value of $F_{\textrm{sp}}$, the distributions of translocation times do not separate by changing the value of $R$ for the region $R \gtrsim R_{\textrm{g}}$ [see panels (f)--(j)]. This results are in agreement with panels (a) and (b) in Fig.~\!\ref{fig_translocation_time}.

\subsection{Waiting time distribution}
\label{subl_wt}

\begin{figure*}[t]
	\begin{minipage}{1.0\textwidth}
    \begin{center}
        \includegraphics[width=0.7\textwidth]{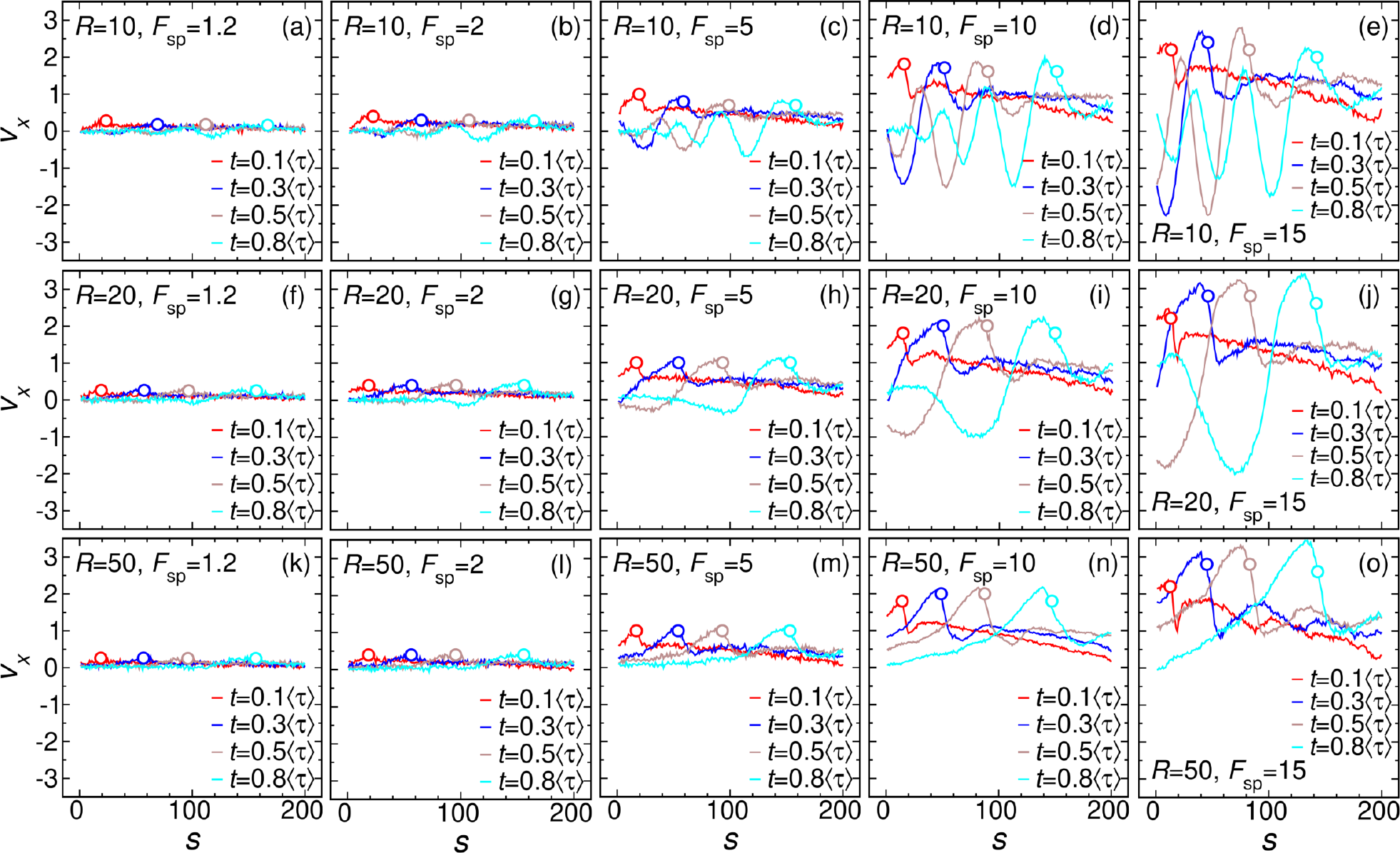}
    \end{center}
    \end{minipage}
\caption{(a) $x$ component of the monomer velocity $v_x$ as a function of the translocation coordinates $s$ for fixed values of $R=10$ and $F_{\textrm{sp}}=1.2$ at different times $t / \langle \tau \rangle =0.1$ (red line), 0.3 (blue line), 0.5 (brown line) and 0.8 (cyan line). 
Panels (b), (c), (d) and (e) are the same as panel (a), but for different values of the SP force $F_{\textrm{sp}}=2$, 5, 10 and 15, respectively. 
Panels (f) to (j), and panels (k) to (o), are the same as panels (a) to (e), but for different values of $R=20$ and 50, respectively. Empty circles show mean value of the index of monomer inside the pore at the corresponding times.}
\label{fig_mvelocity}
\end{figure*}

Waiting time (WT) $w$, which is the time that each monomer spends inside the nano-pore during the translocation process, is an important quantity that reveals the dynamics of the translocation process at the monomer level. 
In Fig.~\!\ref{fig_waiting_time}(a) the WT has been shown as a function of the translocation coordinate $s$ for fixed values of the cavity radius $R=10$ and various values of the SP force $F_{\textrm{sp}}=1.2$ (black line), 2 (red line), 5 (green line), 10 (blue line), and 15 (orange line). Panels (b), (c), and (d), are the same as panel (a), but for different values of the cavity radius $R=12$, 20, and 50, respectively. The translocation coordinate $s$ is the number of monomers at the {\it trans} side. The values of $s=1$ and $s=200$ correspond to the first monomer inside the cavity and the last monomer on the other side of the polymer, respectively. It can be seen that in all panels for the fixed value of $R$ at a given value of $s$, the value of the WT decreases as $F_{\textrm{sp}}$ increases.
As mentioned above, during the equilibration process, the first and the second monomers of the polymer corresponding to $s=1$ and $s=2$, respectively, are inside the cavity and the third monomer with $s=3$ is fixed at the nano-pore. At the beginning of the actual translocation process, when the third monomer is released and the SP force is applied, the first and second monomers have back and forth motions, specially in the weak force limit. Therefore, the value of the  WT for $s=1$ and 2 is not zero and it has a small value (not shown in Fig.~\!\ref{fig_waiting_time}).
The WT has a local maximum at $s=3$, then by increasing $s$ the total net force acting on the {\it trans}-side subchain increases, therefore the value of $w$ decreases. At the end of translocation process $w$ increases as the excluded volume effect grows and resists against the translocation of the monomers into the {\it trans} side. This effect is more pronounced for small values of cavity radius $R=10$ and weak force limit $F_{\textrm{sp}} \lesssim 2$.

Figure \ref{fig_waiting_time}(e) shows the WT in terms of $s$ for a fixed value of SP force $F_{\textrm{sp}}=1.2$ and various values of $R=10$ (black line), 12 (red line), 20 (green line), and 50 (blue line). Panels (f), (g) and (h) are the same as panel (e), but for $F_{\textrm{sp}}=2$, 5, and 10, respectively. 
As the value of the SP force increases [from panel (e) to panel (h)] the curves for different values of the $R$ get closer to each other, and for $F_{\textrm{sp}}=10$ in panel (h) they are almost coincide each other. Consequently, for weak force limit, e.g. for $F_{\textrm{sp}}=1.2$ in panel (e) the value of $R$ plays the dominant role in the dynamics of the translocation process, while in strong force limit, e.g. $F_{\textrm{sp}}=10$ in panel (h), just the SP force controls the translocation dynamics.

\subsection{Monomers' velocities distribution}
\label{subl_mv}

The monomers' velocities in $x$ direction (from {\it cis} to {\it trans}) $v_x$ at time $t$, which is calculated by obtaining the average velocity of each monomer in a narrow time window of $[ t - \delta t /2, t+ \delta t/2 ]$, where $\delta t = 1$, shows the dynamics of the movement of monomers at different times during the translocation process. In Fig.~\!\ref{fig_mvelocity}(a) $v_x$ has been plotted as a function of the translocation coordinates $s$ for fixed values of the cavity radius $R=10$ and the SP force $F_{\textrm{sp}}=1.2$, and at various times $t / \langle \tau \rangle =0.1$ (red line), 0.3 (blue line), 0.5 (brown line) and 0.8 (cyan line). Panels (b), (c), (d) and (e) are the same as panel (a) but for different values of the SP force $F_{\textrm{sp}}=2$, 5, 10 and 15, respectively. 
Panels (f)--(j), and panels (k)--(o) are the same as panels (a)--(e), but for different values of the cavity radius $R=20$ and 50, respectively. Empty circles show mean value of the index of the monomer inside the nano-pore at the corresponding times. These plots indicate that $v_x$ exhibits oscillatory pattern for those monomers inside the cavity. At constant values of $R$ and $F_{\textrm{sp}}$ with the passage of time, more monomers involved in the oscillatory behavior. Moreover, at constant $R$ as the magnitude of $F_{\textrm{sp}}$ increases [from left to right in each row, e.g. from (a) to (e)], the magnitude of the pattern in the oscillatory part of the curves grows. In addition, at constant value of $F_{\textrm{sp}}$ as the value of $R$ increases [from top to bottom in each column, e.g. panels (e)--(j)--(o)] the number of periods decreases, and it is more pronounced at longer times. The oscillations comes from the fact that the polymer subchain in the {\it trans} side (inside the cavity) is folded. If the value of cavity radius is chosen very small, e.g. $R=10$ [panels (a) to (e)], then a spiral is constructed inside the cavity. For constant value of $R=10$, at the long time limit, i.e. $t = 0.8 \langle \tau \rangle$, the oscillations in $v_x$ curve has a damping behavior for those monomers that have entered the cavity first (with small values of $s$) and are located at the inner loops in the spiral due to the lack of enough space. In addition, the monomers that have translocated earlier (with smaller values of $s$) experience weaker net effective force compared to those upcoming monomers at the later times with larger values of $s$. Therefore, the monomers with small values of $s$ have slower dynamics than the monomers with larger values of $s$. This leads to the larger values of the amplitude of $v_x$ for those monomers located at outer layers in the spiral with larger values of $s$ compared to those located at the inner layers with smaller values of $s$. As will be discussed in Sub-sec.~\!\ref{subl_tn}, the strong SP force leads to a relatively perfect spiral inside the cavity with radius $R=10$, while an irregular spiral conformation is made by applying a weak SP force. Another important feature exhibited in Fig.~\!\ref{fig_mvelocity} is that as the value of $R$ increases (from first row in the top to the last row in the bottom) the $v_x$ gets more positive values due to the less excluded volume interactions.

\begin{figure}[t]
    \begin{minipage}{0.5\textwidth}
    \begin{center}
    \hskip-0.32cm
        \includegraphics[width=0.9\textwidth]{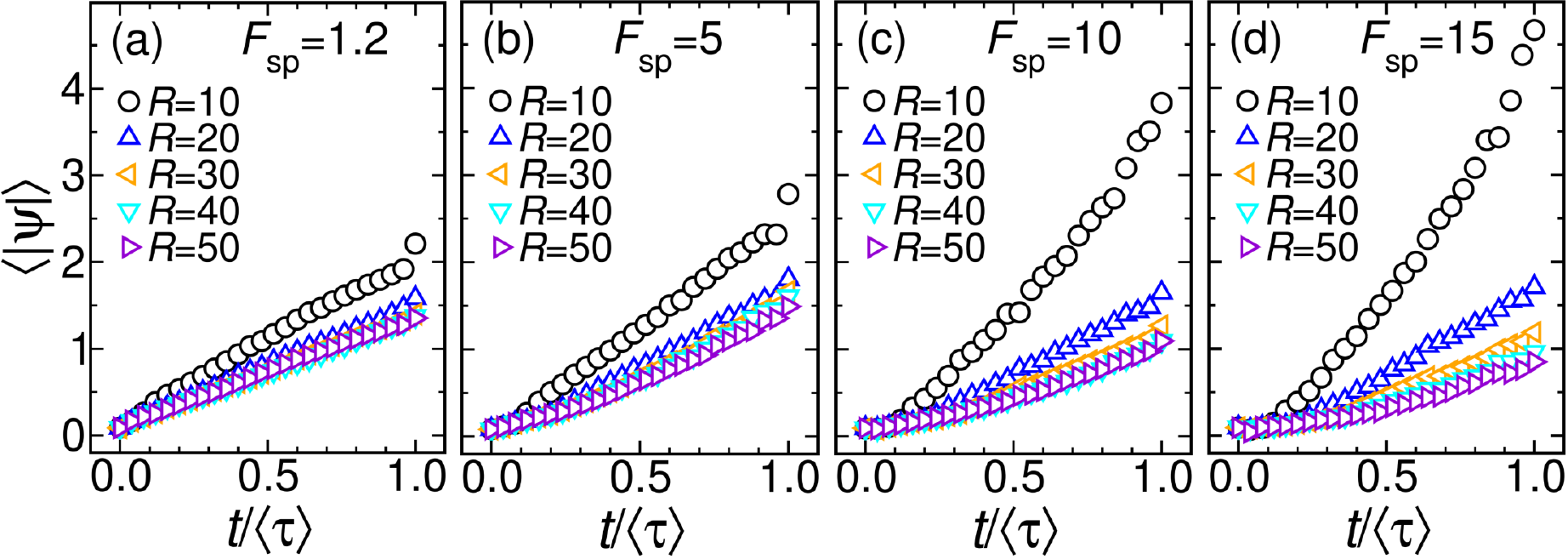}
    \end{center}
    \end{minipage}
\caption{
(a) Average of the magnitude of the turning number $\langle |\psi| \rangle$ as a function of the normalized time $ t / \langle \tau \rangle $ for fixed value of the SP force $F_{\textrm{sp}}=1.2$ and various values of $R=10$ (black circles), 20 (blue triangles up), 30 (orange triangles left), 40 (cyan triangles down) and 50 (violet triangles right). 
Panels (b), (c) and (d) are the same as panel (a) but for different values of $F_{\textrm{sp}}=5$, 10 and 15, respectively.}
\label{fig_turning_number}
\end{figure}

\subsection{Turning number}
\label{subl_tn}

To quantify the properties of the active polymer configurations inside the cavity, the quantity $\psi$, which is the turning number (TN), is introduced as~\!\cite{krantz1999handbook}
\begin{equation}
\psi = \frac{1}{2\pi}\sum_{i=1}^{n}\left[ \theta_{i+1} - \theta_{i} \right],
\label{eq_TN}
\end{equation}
where $i$ and $n$ are the bond index and the total number of bonds inside the cavity at time $t$, respectively, and $\theta_{i}$ is the angle between the $x$ axis and the unit vector $\hat{t_{i}}$ of the bond with index $i$, that is defined by $\hat{t_{i}} = \left( \cos\theta_{i} , \sin\theta_{i} \right)$. 

To study the time evolution of the layered structure of the folded polymer sub-chain inside the cavity, in Fig.~\!\ref{fig_turning_number}(a) the average of the magnitude of the turning number $\langle | \psi | \rangle $ has been plotted as a function of normalized time $t / \langle \tau \rangle $ for fixed value of SP force $F_{\textrm{sp}}=1.2$ and various values of cavity radius $R= 10$ (black circles), 20 (blue triangles up), 30 (orange triangles left), 40 (cyan triangles down) and 50 (violet triangles right). Panels (b), (c) and (d) are the same as panel (a) but for different values of  $F_{\textrm{sp}}=5$, 10 and 15, respectively.
For fixed value of SP force in each panel, as the value of $R$ increases the value of $\langle | \psi | \rangle $ at a given time decreases and approaches to a lower limit value at that moment. Indeed, in each panel at constant value of $F_{\textrm{sp}}$ for $R \gg R_{\textrm{g}}$ the curves almost coincide each other. 
On the other hand for smallest value of the cavity radius $R=10$ and at a constant value of time as the SP force increases [black circles from panel (a) to panel (d)] the value of the $\langle | \psi | \rangle $ increases, that means as the SP force gets stronger a more regular spiral is constructed. In contrast, for $R > R_{\textrm{g}}$ the value of the turning number at a certain time gradually decreases by increasing the value of the SP force.

\begin{figure}[b]
    \begin{minipage}{0.5\textwidth}
    \begin{center}
    \hskip-0.3cm 
        \includegraphics[width=0.96\textwidth]{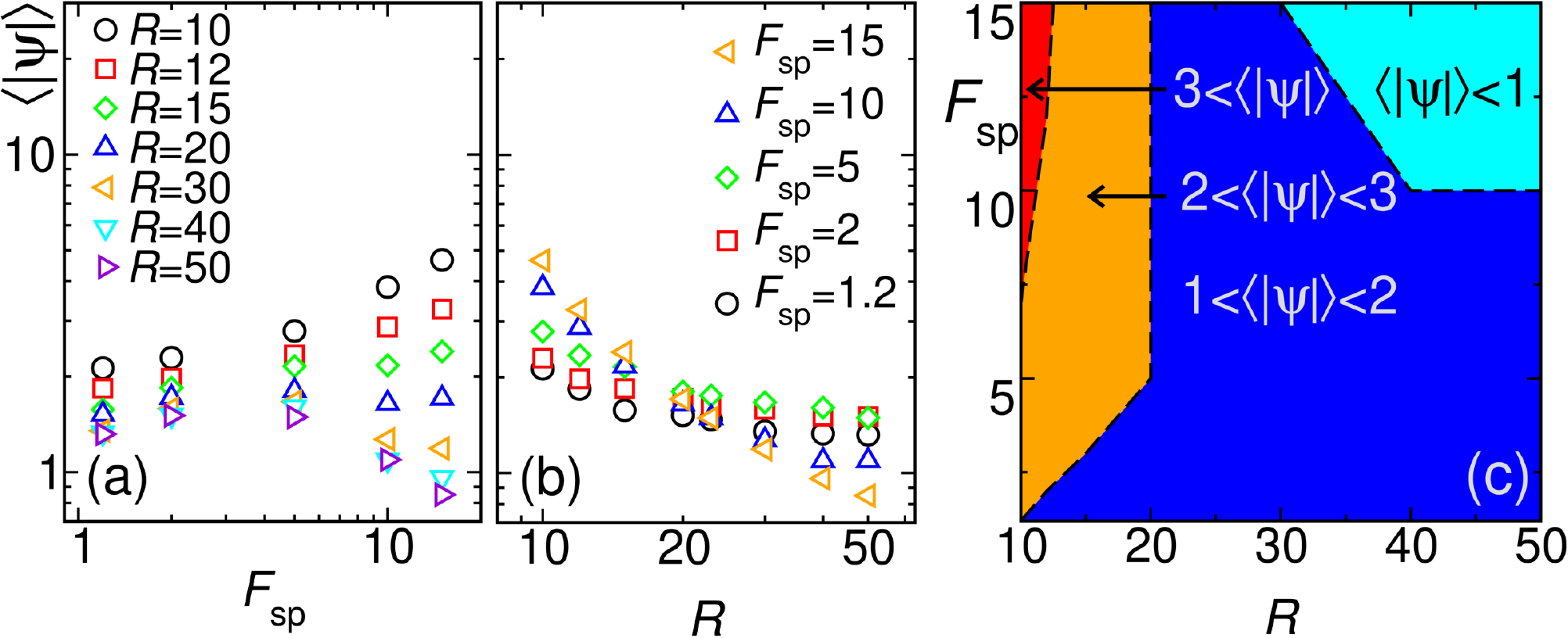}
    \end{center}
    \end{minipage} 
\caption{(a) The magnitude of the turning number $\langle | \psi | \rangle$ for the final snapshot of the system at $t = \tau$, as function of the SP force $F_{\textrm{sp}}$ for various values of the cavity radius $R=10$ (black circles), 12 (red squares), 15 (green diamonds), 20 (blue triangles up), 30 (orange triangles left), 40 (cyan triangles down) and 50 (violet triangles right). 
(b) The $\langle | \psi | \rangle$ as function of $R$ at $t = \tau$, for different values of $F_{\textrm{sp}}=15$ (orange triangles left), 10 (blue triangles up), 5 (green diamonds), 2 (red squares) and 1.2 (black circles).
(c) Phase diagram of $\langle | \psi | \rangle$ at $ t = \tau $ in $R$--$F_{\textrm{sp}}$ plane. The colors cyan, blue, orange and red indicate the values of $\langle | \psi | \rangle < 1$, $1 < \langle | \psi | \rangle < 2$, $2 < \langle | \psi | \rangle < 3$ and $\langle | \psi | \rangle > 3$, respectively.
}
\label{fig_last_turn}
\end{figure}

\begin{figure}[t]
	\begin{minipage}{0.5\textwidth}
    \begin{center}
    \hskip-0.3cm
        \includegraphics[width=0.8\textwidth]{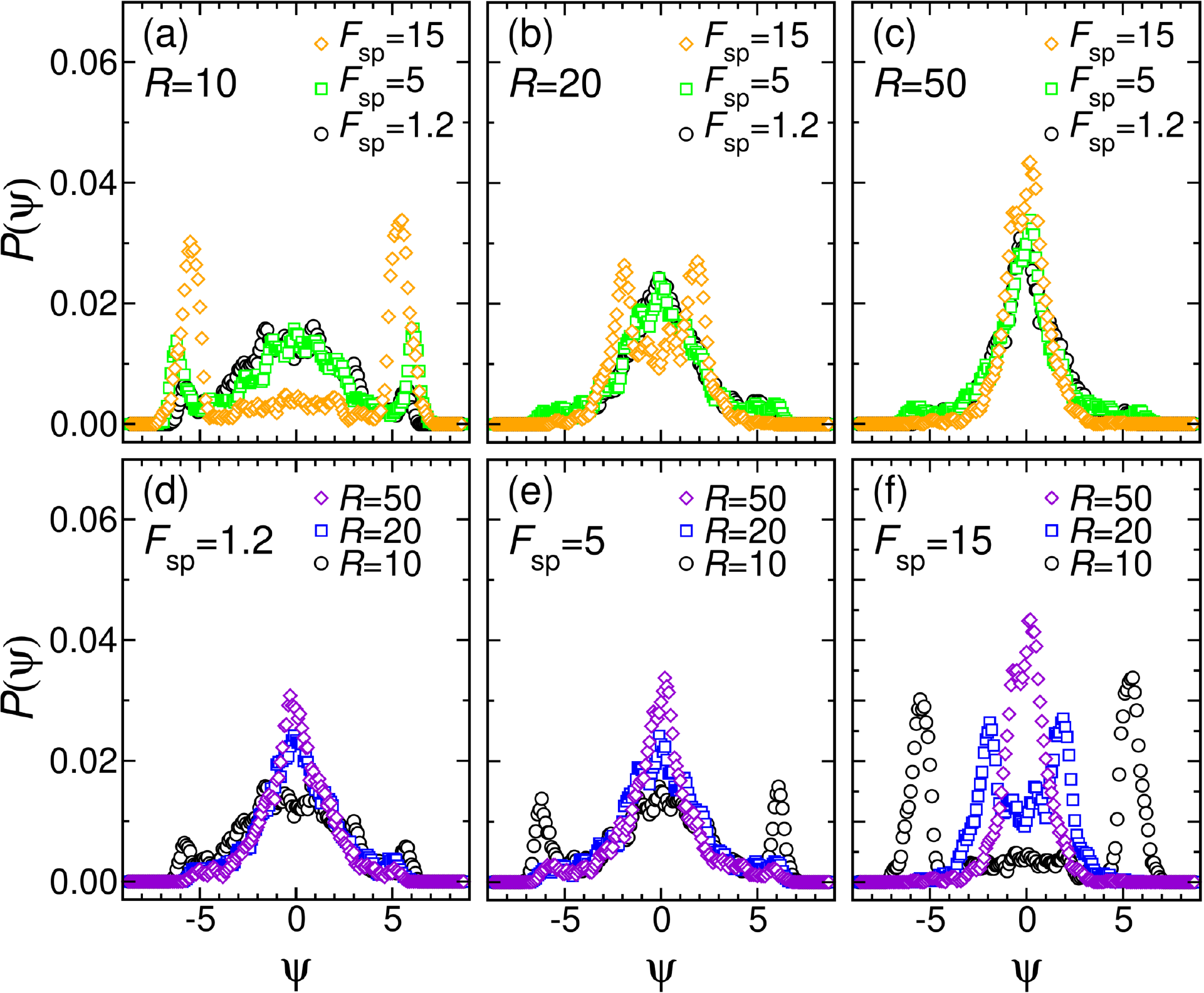}
    \end{center}
    \end{minipage} 
\caption{(a) Probability distribution function of the turning number $P(\psi)$ as a function of turning number $\psi$ for the final snapshot of the system at $t = \tau $, for fixed value of cavity radius $R=10$ and various values of the SP force $F_{\textrm{sp}}=15$ (orange diamonds), 5 (green squares) and 1.2 (black circles). 
Panels (b) and (c) are the same as panel (a) but for different values of $R=20$ and 50, respectively. 
(d) The $P(\psi)$ as a function of $\psi$ at $t = \tau $, for fixed value of $F_{\textrm{sp}}=1.2$ and various values of $R=10$ (black circles), 20 (blue squares) and 50 (violet diamonds). 
Panels (e) and (f) are the same as panel (d) but for different values of SP force $F_{\textrm{sp}}=5$ and 15, respectively.}
\label{fig_distribution_ltn}
\end{figure}

\begin{figure*}[t]
	\begin{minipage}{1\textwidth}
    \begin{center}
    \includegraphics[width=0.75\textwidth]{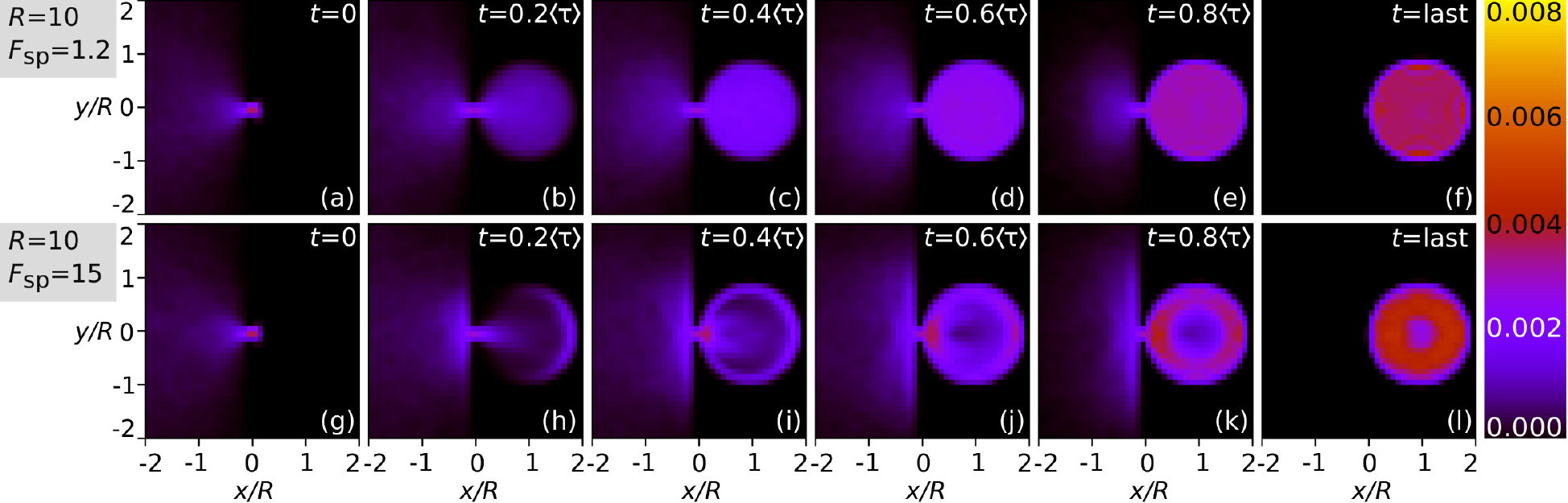}
    \end{center}
    \end{minipage} \hskip-0.0cm      
\caption{
(a) Monomer number density $\rho$ for fixed values of cavity radius $R=10$ and SP force  $F_{\textrm{sp}}=1.2$ at time $t=0$. 
Panels (b), (c), (d), (e) and (f) are the same as panel (a) but at different times $t / \langle \tau \rangle = 0.2$, 0.4, 0.6, 0.8 and the last snapshot, respectively. 
Panels (g)--(l) are the same as panels (a)--(f), respectively, but for $F_{\textrm{sp}}=15$.
}
\label{fig_monomer_density_r10}
\end{figure*}

The value of $\langle | \psi | \rangle $ at the final snapshot of the simulation, when the translocation process terminates, gives information about the structure of the polymer inside the cavity at $t = \tau$. To investigate the polymer structure for the final configuration of the polymer inside the cavity, in Fig.~\!\ref{fig_last_turn}(a) $\langle | \psi | \rangle $ has been plotted as a function of $F_{\textrm{sp}}$ for various values of $R=10$ (black circles), 12 (red squares), 15 (green diamonds), 20 (blue triangles up), 30 (orange triangles left), 40 (cyan triangles down) and 50 (violet triangles right). 
As the value of $F_{\textrm{sp}}$ increases the value of $\langle | \psi | \rangle $ grows monotonically for $ R \lesssim R_{\textrm{g}} $, where $R_{\textrm{g}} \approx 23$. On the other hand, for $ R > R_{\textrm{g}} $, with the increase in the SP force, first the value of $\langle | \psi | \rangle $ grows, then gets its maximum and after that it decreases.

Panel (b) in Fig.~\!\ref{fig_last_turn} shows $\langle | \psi | \rangle $ as a function of $R$ but for various values of SP force $F_{\textrm{sp}}=1.2$ (black circles), 2 (red squares), 5 (green diamonds), 10 (blue triangles up) and 15 (orange triangles left). As seen in the region $ R \lesssim R_{\textrm{g}} $ and at constant $R$ as value of the SP decreases the value of $\langle | \psi | \rangle $ decrease too. On the other hand, in the region $ R > R_{\textrm{g}} $ and at constant $R$, for $F_{\textrm{sp}} \geq 5$ as the value of the SP force increases the value of $\langle | \psi | \rangle $ decreases, and for $F_{\textrm{sp}} \leq 2$ it's vice versa.

In panel (c) in Fig.~\!\ref{fig_last_turn} the results of panels (a) and (b) have been qualitatively summarized by drawing a phase diagram for quantity $\langle | \psi | \rangle $ for the final snapshot of the polymer configuration at $ t = \tau $ in the phase space $R$--$F_{\textrm{sp}}$. According to the values of $\langle | \psi | \rangle $ for different sets of $R$--$F_{\textrm{sp}}$, the phase space has been divided into four regions $\langle | \psi | \rangle  < 1$ (in cyan color), $ 1 < \langle | \psi | \rangle < 2$ (in blue color),  $ 2 < \langle | \psi | \rangle < 3$ (in orange color) and  $ \langle | \psi | \rangle > 3$ (in red color).  

To investigate the turning number $\psi$ in more detail, in Fig.~\!\ref{fig_distribution_ltn}(a) the probability distribution function of the turning number $P(\psi)$ has been plotted in terms of $\psi$ at $t = \tau$ for the whole active polymer inside the cavity for fixed value of $R=10$ and various values of the SP force $F_{\textrm{sp}}=1.2$ (black circles), 5 (green squares) and 15 (orange diamonds). Panels (b) and (c) are the same as panel (a) but for different values of $R=20$ and 50, respectively.
Panel (d) shows $P(\psi)$ as a function of $\psi$ for fixed value of $F_{\textrm{sp}}=1.2$ and various values of cavity radius $R=10$ (black circles), 20 (blue squares) and 50 (violet diamonds). Panels (e) and (f) are the same as panel (d), but for different values of $F_{\textrm{sp}}=5$ and 15, respectively.

According to the horizontal axis of the panels in Fig.~\!\ref{fig_distribution_ltn}, the value of $\psi$ can be negative (positive) that shows the clockwise (counter clockwise) conformation of a spiral. As can be seen from panel (a), the translocation of an active polymer for $F_{\textrm{sp}}=15$ into a cavity with radius $R=10$, leads to two peaks at $\psi \approx -5$ and 5 in $P (\psi)$. In the region between the two peaks the value of $P(\psi)$ are very small compared to its values at the peaks. Infact for the smallest chosen value of the cavity radius $R=10$ and strong SP force $F_{\textrm{sp}}=15$, the active polymer folds to make mostly a conformation with regular spiral structure with $\psi \approx -5$ ($+5$) as a clockwise (counter clockwise). For smaller values of the SP force, the height of the peaks decreases, and instead a hump in the region between the side peaks appears with a maximum around $\psi=0$. Indeed, for $R=10$ by decreasing the value of SP force the translocation process takes place in such a way that the polymer folds inside the cavity and finally at $t=\tau$ gets an irregular spiral conformation. This result is in agreement with the results of previous subsections. 

\begin{figure*}[t]
	\begin{minipage}{1\textwidth}
    \begin{center}
    \includegraphics[width=0.75\textwidth]{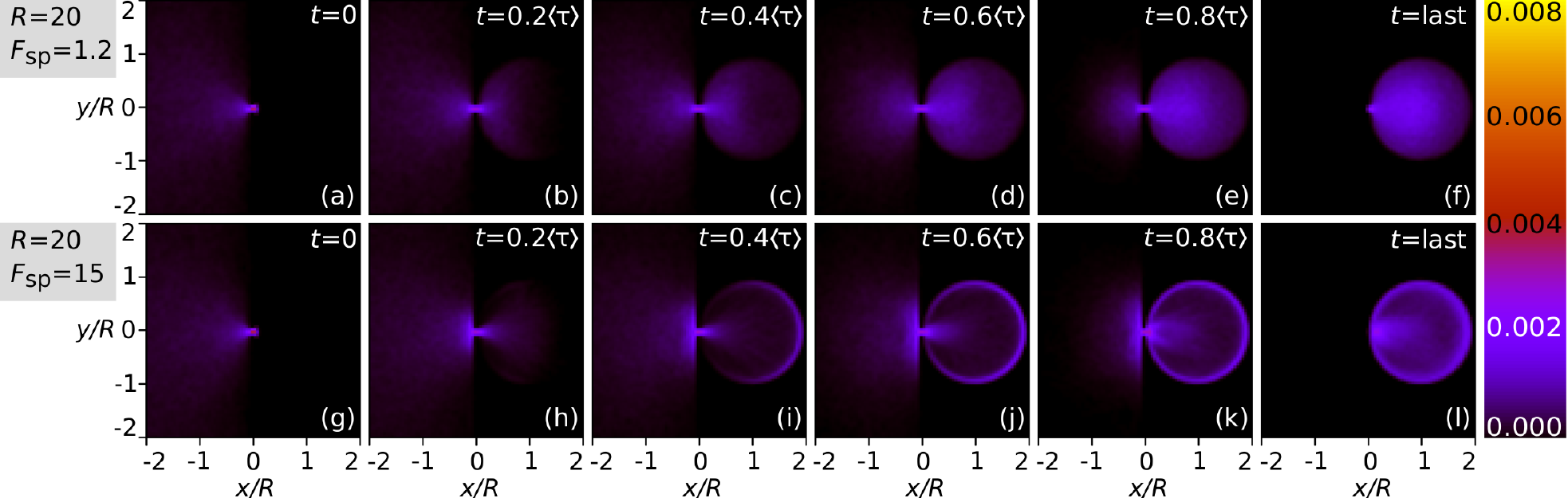}
    \end{center}
    \end{minipage} \hskip-0.0cm      
\caption{
(a) Monomer number density $\rho$ for fixed values of cavity radius $R=20$ and SP force $F_{\textrm{sp}}=1.2$ at time $t=0$. 
Panels (b), (c), (d), (e) and (f) are the same as panel (a) but at different times $t / \langle \tau \rangle = 0.2$, 0.4, 0.6, 0.8 and the last snapshot, respectively. 
Panels (g)--(l) are the same as panels (a)--(f), respectively, but for $F_{\textrm{sp}}=15$.
}
\label{fig_monomer_density_r20}
\end{figure*}

\begin{figure*}[t]
	\begin{minipage}{1\textwidth}
    \begin{center}
    \includegraphics[width=0.9\textwidth]{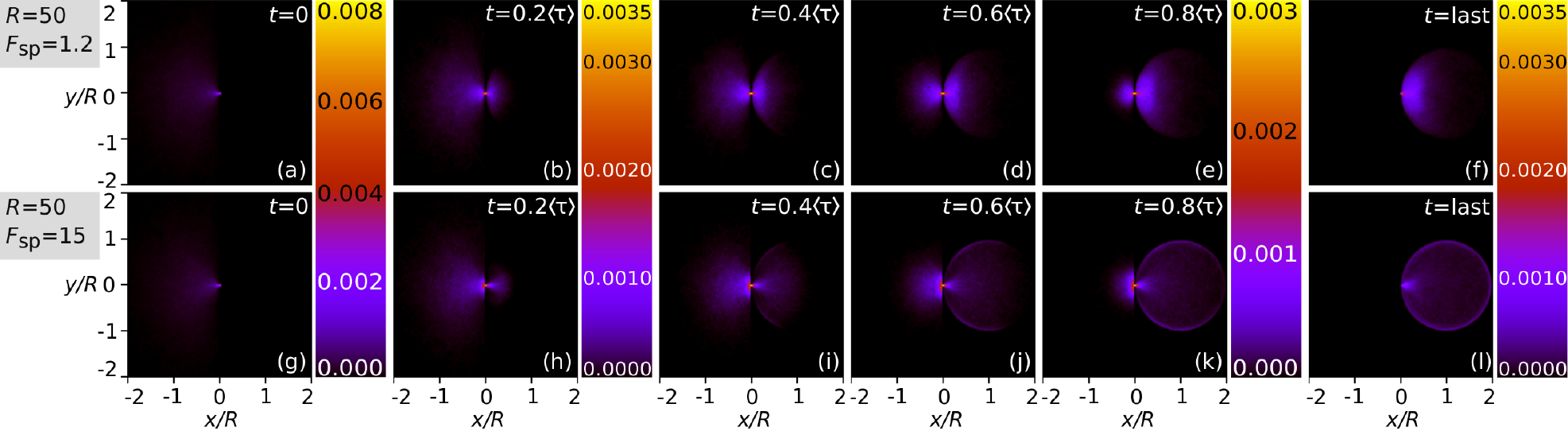}
    \end{center}
    \end{minipage} \hskip-0.0cm     
\caption{
(a) Monomer number density $\rho$ for fixed values of cavity radius $R=50$ and SP force  $F_{\textrm{sp}}=1.2$ at time $t=0$. 
Panels (b), (c), (d), (e) and (f) are the same as panel (a) but at different times $t / \langle \tau \rangle = 0.2$, 0.4, 0.6, 0.8 and the last snapshot, respectively. 
Panels (g)--(l) are the same as panels (a)--(f), respectively, but for $F_{\textrm{sp}}=15$.
}
\label{fig_monomer_density_r50}
\end{figure*}


For the strong SP force $F_{\textrm{sp}}=15$ (orange diamonds) as the value of $R$ increases from panel (a) to (c) the height of the side peaks in the distribution decreases and their locations get closer to each other. and finally in panel (c) the two side peaks are merged. 
Moreover, for smaller values of the SP force by increasing the value of $R$ the height of the side peaks gets smaller and the hump at the middle of the distribution is more pronounced [from panel (a) to (c)] and the dominant part of the distribution is in the middle hump.

Indeed, for $R=50$ in panel (c) the decrease in the value of the SP force leads to a relatively wider distribution. This confirms that by increasing the size of the circular cavity, the twine picks and therefore the regular spiral structure of the active polymer inside the cavity disappear [see panels (b) and (c) in Fig.~\!\ref{fig_last_turn}].


\subsection{Monomer density}
\label{subl_md}

The time evolution of the spatial configuration of the active polymer during the translocation process can be considered by the monomers number density $\rho$. To quantify the $\rho$, the simulation box is divided into square cells each with size of $1 \times 1$. Then the average numbers of monomers that occupy the cells are obtained in a narrow window of time $\Delta t =1$ around a desired moment for each trajectories and after that the average is taken over all trajectories at the corresponding time as well as over the narrow window of time. The density for the final time step of the translocation process is obtained only by taking the average over the trajectories. 

Fig.~\!\ref{fig_monomer_density_r10} demonstrates the time evolution of $\rho$. Panel (a) shows $\rho$ just at the beginning of the translocation process at $t=0$ for fixed values of $R=10$ and $F_{\textrm{sp}} = 1.2$. Panels (b), (c), (d), (e) and (f) are the same as panel (a) but at times $t=0.2 \langle \tau \rangle $, $0.4 \langle \tau \rangle $, $0.6 \langle \tau \rangle $, $0.8 \langle \tau \rangle $ and at the final time step of the translocation process, respectively. As seen in panel (a) at time $t=0$ 
the monomers are symmetrically distributed around the horizontal axis passing through the center of the nano-pore. Panels (b) to (f) show that for small cavity with $R=10$ at a weak force limit $F_{\textrm{sp}}=1.2$ the monomers are almost uniformly distributed inside the cavity during the translocation process.
Panels (g)--(l) are the same as panels (a)--(f) but for strong value of $F_{\textrm{sp}}=15$.  Panel (h), which presents $\rho$ at $t = 0.2 \langle \tau \rangle$, clearly shows that for strong value of SP force the dynamics is very fast and the sub-chain inside the cavity does not have enough time to fold itself. Therefore, the polymer is able to touch the cavity wall in front of the nano-pore in the other side of the cavity at short time limit. With the passage of the time, as presented in panels (i) to (l), the sub-chain inside the cavity folds due to the interaction with the cavity wall. This leads to a ring structure in vicinity of the cavity wall as seen at $t = 0.4 \langle \tau \rangle$ in panel (i) and at the later times a regular structure which has a spiral configuration is being made as illustrated in panels (j) to (l).

In Fig.~\!\ref{fig_monomer_density_r10}, the results of panels (a)--(f) and panels (g)--(l) for weak and strong SP force limits, respectively, are in agreement with the results shown in Fig.~\!\ref{fig_distribution_ltn}(a). Indeed for weak force limit of $F_{\textrm{sp}}=1.2$ while the polymer is translocating it has enough time to fold itself. Therefore, almost at all times even at the final time step of the translocation process [panel(f)] the density is uniform. This is the feature of irregular configurations of the polymer subchain inside the cavity at all times. This leads to the wide distribution of the turning number for $R=10$ (black circles) in Fig.~\!\ref{fig_distribution_ltn}(a). In contrast, for strong force $F_{\textrm{sp}}=15$ the polymer dynamics is fast and it touches the front side of the cavity wall at short time and the ring-shaped at the vicinity of the cavity wall is constructed and as time passes this structure becomes thicker [red color region in panel (l)]. Formation of the ring-shaped structure for the monomer density in panel (l) is completely compatible with the presented result in Fig.~\!\ref{fig_distribution_ltn}(a) for $R=10$ (orange diamonds), in which the probability distribution of the turning number has two peaks at $\psi \approx 5$ and -5 indicating that the active polymer with strong SP force has a regular structure at the end of the translocation process.
This behavior refers to the fact that, formation of the spiral structure at the strong SP force regime $F_{\textrm{sp}}=15$ inside a circular cavity with radius $R=10$, mostly, is from outer layer to the inner one. On the other hand, in the weak SP force regime $F_{\textrm{sp}}=1.2$ during the folding, the polymer inside the cavity has enough time to explore different spatial configurational states due to the thermal fluctuations, therefore, the polymer is irregularly packed.

Figures~\!\ref{fig_monomer_density_r20} and \ref{fig_monomer_density_r50} are the same as Fig.~\!\ref{fig_monomer_density_r10} but for different values of cavity radius $R=20$ and 50, respectively. Comparing Figs.~\!\ref{fig_monomer_density_r20} and \ref{fig_monomer_density_r50} with Fig.~\!\ref{fig_monomer_density_r10} reveals that by increasing the value of $R$ those effects discussed above for small cavity radius $R=10$ 
are become less pronounced. 

At weak force regime of $F_{\textrm{sp}}=1.2$, as the value of $R$ increases the available space for the polymer sub-chain inside the cavity increases too. This leads to funnel shape for the monomer density at short time limit [see panel (b) in Fig.~\!\ref{fig_monomer_density_r20}]. With the passage of the time the size of the funnel grows and at the end of translocation process [panel (f) in Fig.~\!\ref{fig_monomer_density_r20}] changes to a pear shape  with the larger density in the vicinity of the nano-pore and smaller density close to the cavity wall specially next to cavity wall in front of the nano-pore.

In the limit of strong SP force $F_{\textrm{sp}}=15$ with $R=20$ at short time limit, e.g. $t=0.2 \langle \tau \rangle$, as depicted in panel (h) of Fig.~\!\ref{fig_monomer_density_r20} similar to panel (b) a funnel shape is made. But in contrast to panels (c) to (f), two separate regions with higher values of the monomer density evolve as seen in panels (i) to (l). The first high density region is in the vicinity of the nano-pore in which due to the high value of SP force the monomers are accumulated in. The other high density region is the vicinity of the cavity wall wherein the polymer sub-chain interacts with and spend more time there. The results for the distribution of turning number (blue squares) in panels (d) and (f) in Fig.~\!\ref{fig_distribution_ltn} confirm that the configuration of the polymer inside the cavity with radius $R=20$ at the end of translocation process is more regular for the strong force $F_{\textrm{sp}}=15$  than that of the weak force $F_{\textrm{sp}}=1.2$.

The results in Fig.~\!\ref{fig_monomer_density_r50} are similar to those of Fig.~\!\ref{fig_monomer_density_r20} but are less pronounced due to the fact that there is more available space for $R=50$ as mentioned above. Moreover, as seen in Fig.~\!\ref{fig_monomer_density_r50} the large value of the available space leads to the irregular configurations at both weak and strong force limits due to the lack of the spiral configurations, which is in agreement with the results for the distribution of turning number (violet diamonds) in panels (d) and (f) in Fig.~\!\ref{fig_distribution_ltn}. 


\section{Summary and conclusions}
\label{secl_SC}

In summary we have investigated the dynamics of an active semi-flexible polymer translocation into a circular cavity with the radius of $R$ in two dimension using the Langevin dynamics simulation method. An active polymer has been modeled by applying the SP force $F_{\textrm{sp}}$ on the specified segments of the polymer. 
In addition to the force exponent the detail dynamics of the process has been explored by the study of different quantities such as waiting time distribution, velocities of monomers, turning number and monomer density.

The most important quantity that reveals the global dynamics of translocation process is the translocation time $\tau$. The relation between $\tau$ and the SP force $F_{\textrm{sp}}$ introduces the force exponent $\beta$ as $\tau \propto F_{\textrm{sp}}^{\beta}$. Our results show that the force exponent $\beta$ has a non-monotonic behavior with respect to the cavity radius. For the smallest value of $R=10$ in the present study, the value of the force exponent is $\beta \approx -1$ and by increasing the value of $R$ the value of $\beta$ grows and it gets its maximum at $R \approx R_{\textrm{g}}$, where $ R_{\textrm{g}} \approx 23$ is the average value of the radius of gyration for a passive semi-flexible polymer in the free space in two dimension. For large values of $R$, the $\beta$ approaches its asymptotic value $\beta \approx -0.92$.

\begin{figure*}[t]
	\begin{minipage}{1.0\textwidth}
    \begin{center}
        \includegraphics[width=0.7\textwidth]{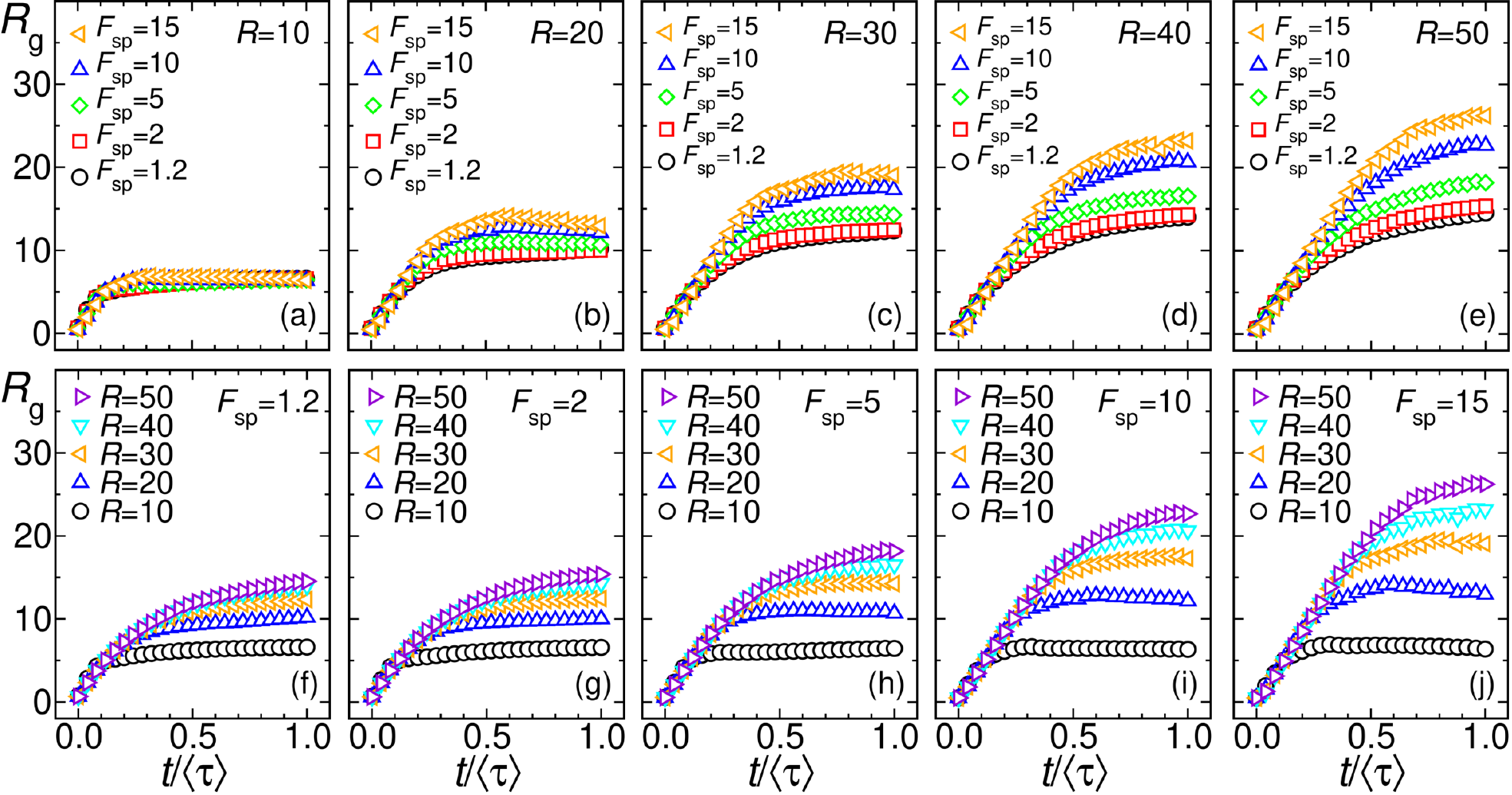}
    \end{center}
    \end{minipage} 
\caption{(a) The average radius of gyration $R_{\textrm{g}}$ as a function of the normalized time $t / \langle \tau \rangle$ for fixed value of the cavity radius $R=10$ and various values of the SP force $F_{\textrm{sp}}=15$ (orange triangles left), 10 (blue triangles up), 5 (green diamonds), 2 (red squares) and 1.2 (black circles). 
Panels (b), (c), (d) and (e) are the same as panel (a) but for different values of $R=20$, 30, 40 and 50, respectively.
(f) The $R_{\textrm{g}}$ as a function of $t / \langle \tau \rangle$ for fixed value of $F_{\textrm{sp}}=1.2$ and various values of $R=10$ (black circles), 12 (red squares), 15 (green diamonds), 20 (blue triangles up), 30 (orange triangles left), 40 (cyan triangles down) and 50 (violet triangles right). 
Panels (g), (h), (i) and (j) are the same as panel (f) but for different values of $F_{\textrm{sp}}=2$, 5, 10 and 15, respectively.}
\label{fig_rg}
\end{figure*}

The waiting time distribution, which is the time that each monomer spends at the nano-pore during the translocation process, illustrates the dynamics of the process at the monomer level. The results of the present study show that at constant value of the cavity radius, the waiting time decreases by increasing the value of the SP force in both the weak as well as the strong force limits [panels (a)--(d) in Fig.~\!\ref{fig_waiting_time}]. On the other hand, at constant value of SP force and at weak force limit the waiting time is sensitive to the value of $R$ [see panel (e) in Fig.~\!\ref{fig_waiting_time}], while at the strong SP force limit [panel (h) in Fig.~\!\ref{fig_waiting_time}] the value of $w$ does not depend on the value of $R$ as the main contribution to the dynamics of the translocation process comes from the SP force.  

The next quantity that has been considered is the $x$ component of the monomer velocities as a function of the $s$ at different times during the translocation process. At constant values of $R$ and $F_{\textrm{sp}}$ as time passes the oscillations in the velocity curves appear [different rows, e.g. panels (a)-(e), in Fig.~\!\ref{fig_mvelocity}]. The oscillations in the velocity curves become more visible as the value of the SP force increases [e.g. panels (a) to (e) in Fig.~\!\ref{fig_mvelocity}]. As the value of the cavity radius increases the wavelength of the oscillations increases and the number of periods of the oscillations decreases [see e.g. panels (c)-(h)-(m) or panels (e)-(j)-(o) in Fig.~\!\ref{fig_mvelocity}]. 

The above results indicate that as the value of $R$ decreases, the configuration of the polymer inside the cavity become in average more regular. To quantify this, the turning number $\psi$, which is the average of the number of turns in the polymer configuration [see Eq.~\!\ref{eq_TN}] has been studied. As seen in Fig.~\!\ref{fig_last_turn}(c) for small values of $R$ and strong force the high value of $\langle |\psi| \rangle$ confirms that the polymer configuration is more regular and makes the spiral form. In contrast, at large value of $R$ and small value of the SP force, the polymer mostly has an irregular configuration.

To see the dynamics of the average spatial configuration of the polymer during the translocation process the monomer number density has been investigated at different times. For smallest value of cavity radius $R=10$ and the weak SP force $F_{\textrm{sp}} = 1.2$ the monomers are uniformly distributed inside the cavity almost at all moments during the translocation process [except at the beginning of the process, panel (b)--(f) in Fig.~\!\ref{fig_monomer_density_r10}]. At the strong SP force limit of $F_{\textrm{sp}} = 15$ a ring structure is constructed and with the passage of the time the thickness of the ring increases. This confirms that the polymer configuration in average makes a regular spiral [panels (h)--(l) in Fig.~\!\ref{fig_monomer_density_r10}].
With the increase of cavity radius to $R=20$ and at the weak force limit of $F_{\textrm{sp}} = 1.2$, the density makes a pear shape and as time passes the pear occupies the whole cavity space [panels (b)--(f) in Fig.~\!\ref{fig_monomer_density_r20}]. At constant $R=20$ for the strong force $F_{\textrm{sp}} = 15$ the density is more pronounced in two regions in the cavity. The first region is the vicinity of the nano-pore and the second one is next to the cavity wall in front of the nano-pore [e.g. panel (i) in Fig.~\!\ref{fig_monomer_density_r20}]. With the passage of the time both regions extend [panels (h)--(l) in Fig.~\!\ref{fig_monomer_density_r20}]. Especially at the end of the translocation process the second region extend around the whole cavity wall [panel (l) in Fig.~\!\ref{fig_monomer_density_r20}].    
With the increase in the cavity radius even further to $R=50$, the less pronounced results are obtained compared to those for $R=20$. 

The results of the present study shed light on the deeper understanding about the dynamics of the injection of an active polymers into a confined space such as biological cells.


\begin{acknowledgments}
A.R.-D. is indebted to Dr. Farshid Mohammad-Rafiee for his compassionate supports.
\end{acknowledgments}


\appendix


\section{Radius of gyration}
\label{subl_rg}

The time evolution of the radius of gyration (RG) is presented in this appendix. The RG for a polymer configuration is obtained by calculating the average distances of the monomers from the center of mass of the polymer. Then, averaging over different trajectories gives a general information about the average spatial size of the polymer. The explicit relation to calculate the average of the RG is given by~\!\cite{RubinsteinBook} 
\begin{equation}
R^{2}_{\textrm{g}} = \frac{1}{N} \sum_{i=1}^{N} \langle (\vec{r}_{i} - \vec{r}_{\textrm{CM}})^{2} \rangle,
\label{eq_rg}
\end{equation}
where, $R_{\textrm{g}}$ is the average value of the RG, $\vec{r}_{i}$ and $\vec{r}_{\textrm{CM}} = 1/N \sum_{i=1}^{N} \vec{r}_{i}$ are position vector of the {\it i}th monomer and center of mass vector of the polymer, respectively, with {\it i} as the index of the monomers and $N$ is the total number of monomers in the polymer. Here, $\langle \ldots \rangle $ denotes the ensemble average over different trajectories.  To unveil the time evolution of RG, in Fig.~\!\ref{fig_rg}(a) $R_{\textrm{g}}$ of the {\it trans}-side subchain has been plotted as a function of the normalized time $t / \langle \tau \rangle $ for fixed value of the cavity radius $R=10$ and for various values of the SP forces $F_{\textrm{sp}}=1.2$ (black circles), 2 (red squares), 5 (green diamonds), 10 (blue triangles up) and 15 (orange triangles left). Panels (b), (c), (d) and (e) are the same as panel (a) but for different values of $R=20$, 30, 40 and 50, respectively. 
In panel (a), the curves for different values of $F_{\textrm{sp}}$ are almost coincide each other, due to the small value of $R=10$ that leads to the small available spatial space to the polymer. As the value of the cavity radius increases, from panel (a) to (e), more spatial space is available to the {\it trans}-side polymer subchain. Therefore, as the value of the SP force increases the {\it trans}-side polymer subchain explores more space that yields the larger value of the RG at longer time scales.
Panel (f) demonstrates $R_{\textrm{g}}$ as a function of $t / \langle \tau \rangle $ for fixed value of SP force $F_{\textrm{sp}}=1.2$ and various values of $R=10$ (black circles), 20 (blue triangles up), 30 (orange triangles left), 40 (cyan triangles down) and 50 (violet triangles right). As can be seen by increasing the value of $F_{\textrm{sp}}$, from panel (f) to panel (j), the curves deviate more from each other. It means that for strong SP force [panel (j)] with increasing the value of $R$, monomers explore more space, than that of the weak SP force in panel (f).




\end{document}